# Robust micro-magnet design for fast electrical manipulations of single spins in quantum dots

Jun Yoneda[1,2,†], Tomohiro Otsuka[1,2], Tatsuki Takakura[1], Michel Pioro-Ladrière[3,4], Roland Brunner[5], Hong Lu[6,*], Takashi Nakajima[1,2], Toshiaki Obata[1], Akito Noiri[1], Christopher J. Palmstrøm[6], Arthur C. Gossard[6], and Seigo Tarucha[1,2]

[1] *Department of Applied Physics, University of Tokyo, Tokyo 113-8656, Japan*
[2] *RIKEN Center for Emergent Matter Science, RIKEN, 2-1 Hirosawa, Wako, Saitama 351-0198, Japan*
[3] *Départment de Physique, Université de Sherbrooke, Quebec J1K 2R1, Canada*
[4] *CIFAR Program in Quantum Information Science, Canadian Institute for Advanced Research (CIFAR), Toronto, Ontario M5G 1Z8, Canada*
[5] *Department of Materials for Microelectronics, Materials Center Leoben Forschung GmbH, 8700, Austria*
[6] *Materials Department and Department of Electrical and Computer Engineering, University of California, Santa Barbara, CA 93106, USA*

†Author to whom correspondence should be addressed.    Email: jun.yoneda@riken.jp
*Currently at College of Engineering and Applied Sciences, Nanjing University, Nanjing, Jiangsu 210093, China

**Abstract.** Tailoring spin coupling to electric fields is central to spintronics and spin-based quantum information processing. We present an optimal micromagnet design that produces appropriate stray magnetic fields to mediate fast electrical spin manipulations in nanodevices. We quantify the practical requirements for spatial field inhomogeneity and tolerance for misalignment with spins, and propose a design scheme to improve the spin-rotation frequency (to exceed 50MHz in GaAs nanostructures). We then validate our design by experiments in separate devices. Our results will open a route to rapidly control solid-state electron spins with limited lifetimes and to study coherent spin dynamics in solids.

Electronics based on spin degrees of freedom, *e.g.* spintronics and spin-based quantum computing [1,2], has attracted much attention recently as approaches to overcome the scaling limit of conventional semiconductor microelectronics. Such spin-based processing requires the ability to rapidly manipulate electron spins before spin information is lost. A canonical method to control spins in solid-state nanoscale devices is electron spin resonance (ESR) [1-3]. Recent progress in the nanoelectronics science has led to observations of ESR at the single electron level by several different mediating mechanisms in semiconductor quantum dot (QD) devices [4-7]. However, control strengths are often bounded by material properties and weaker than noises in the condensed-matter environment, which destroy the quantum nature of spin during manipulation. For coherent spin rotations with no unintentional phase shift, one needs fast ESR (> 50 MHz e.g. for GaAs [8]), which acts on timescales shorter than the dephasing time (~ 40 ns e.g. in GaAs). As two-spin entanglement can be gated on much faster timescales [9,10], a generic approach to enable fast ESR would provide means to high-fidelity universal operations of spin qubits in a quantum processor [3] and control of the coherent superposition state of solid-state electron spins possibly interacting relatively strongly with the environment.

In this letter, we focus on the electrically driven ESR in QD devices with micro-magnets (MMs). The integration of MMs gives the following advantages for local control of single spins [7,8,10,11-14]. First, ESR can be electrically driven in a slanting magnetic field induced by MMs (MM-ESR) [11,7]. For practical applications, control by electric fields is particularly appealing, since they are more readily generated locally than magnetic fields. MM-ESR has shown to generate the fastest electrical spin rotations so far in semiconductor QDs [8] and can also boost ESR speed mediated by the spin-orbit interaction [15,16,5]. Second, the technique is highly scalable and can be extended to more than 25 spin-qubits [15]. Third, it is material independent [11,13], and applicable to QDs fabricated in various materials, *e.g.* isotopically-purified C- or Si-based semiconductors with longer spin coherence times. However, for the application of the MM approach to fast spin control, a careful design of the MM is mandatory. In what follows, we show comprehensively how the local magnetic field produced by MMs can be tailored to facilitate fast, addressable ESR necessary for QD-spin-based spintronics and quantum information processing.

To explore the possibility of high-fidelity spin rotations with the MM, it is of primary importance to clarify the required conditions of its local field. In MM-ESR, two properties of the magnetic field are mainly utilized: the linear gradient of the magnetic field component normal to the spin quantization axis ($b_{sl}$) and the difference in the Zeeman field between QDs

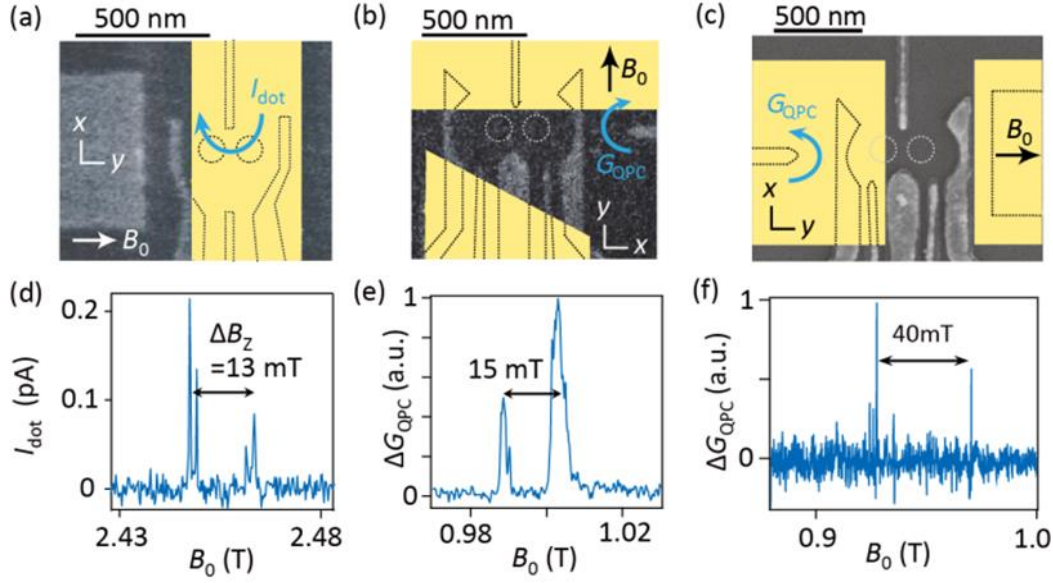

Figure 1. **(a)-(c)** Scanning electron micrographs of GaAs DQD devices with false color Co MMs on top: (a) Ref. [7], (b) Ref. [10], and (c) Ref. [17]. The DQD is defined electrostatically by negatively biased Schottky metal gates deposited on a modulation-doped GaAs/AlGaAs wafer with a 2DEG 90 nm below the surface. A 70-160 nm thick Co ferromagnet is deposited on top of ~100 nm thick insulating Calixarene and is magnetized under a sufficiently large external magnetic field in the plane of the 2DEG ($B_0 \gtrsim 1$ T). Spins are flipped under a.c. gate-voltage excitation when the excitation frequency matches their Larmor frequency. Lift of spin blockade by MM-ESR is detected by measuring either a current through the DQD ((a)) or the change in conductance of an adjacent quantum point contact ((b)-(c)), while the DQD is tuned in the Pauli spin-blockade regime [20,21]. **(d)-(f)** ESR spectra measured in these devices. [(a) and (d) are reprinted by permission from Macmillan Publishers Ltd: *Nature Physics* [7], copyright 2008.]

($\Delta B_Z$). Here, the Zeeman field is defined as the magnetic field component parallel to the spin quantization axis. As formulated in Ref. [11], $b_{sl}$ couples the electron's spin and orbital degrees of freedom and allows electrically driven ESR. $\Delta B_Z$, on the other hand, gives different Larmor frequencies for spins in different QDs, allowing access to a single spin without affecting others. To quantify the required field strengths, we first analyze the conventional MM designs which have been employed previously in GaAs double QD (DQD) devices [7,10,17] (Fig. 1). For each device two ESR peaks were observed, which demonstrates the control addressability. From the ESR peak separation, $\Delta B_Z$ can be measured directly. On the other hand, the exact values of $b_{sl}$ are difficult to extract from the experimental data. Although the ESR peak height reflects the Rabi frequency $f_{Rabi}$, $f_{Rabi}$ is not solely determined by the value of $b_{sl}$, but proportional to the product of both $b_{sl}$ and the amplitude of the a.c. driving electric field $E_{ac}$ to leading order of $b_{sl}$ and $E_{ac}$ [18]. Therefore, we numerically simulate the MM properties [19], $b_{sl}$ and $\Delta B_Z$, using a boundary integral method (Table I). In the simulation, we take the device-dependent geometrical parameters into account, which may modify the MM properties. We also allow for 75 nm misalignment of the MM pattern with respect to the QDs that is possibly present in the real devices. Other sources of parameter error *e.g.* the interdot-distance estimation can also influence but the effect is not significant. The consistency of the simulated MM properties and experimental observation (Table I) supports the validity of this simulation. All MMs satisfy the conditions $b_{sl} \gtrsim 0.4$ mT/nm and $\Delta B_Z \gtrsim 10$ mT, which we interpret as a sufficient condition to discern individual ESR peaks in GaAs QDs.

TABLE I. Simulated and observed MM field properties in previous MM-ESR experiments along with geometrical parameters

|  | Simulation (max, min) [a] | | | Experiment | | Parameter | |
|---|---|---|---|---|---|---|---|
|  | Larger $b_{sl}$ [mT/nm] | Smaller $b_{sl}$ [mT/nm] | $\Delta B_Z$ [mT] | Estimated $b_{sl}$ [mT/nm] | $\Delta B_Z$ [mT] | $t_{MM}$[b] [nm] | $d_{2DEG}$[c] [nm] |
| Ref. (a) | 0.51, 0.46 | 0.50, 0.0 | 61, -9.2 | 0.8 | 13 ± 2 | 70 | 170 |
| Ref. (b) | 0.54, 0.36 | 0.49, 0.33 | 8.6, 5.1 | - | 15 ± 5 | 150 | 210 |
| Ref. (c) | 0.62, 0.48 | 0.48, -0.15 | 84, 31 | - | 40 ± 5 | 160 | 210 |
| Device A | 1.53, 1.11 | 1.46, 0.90 | 48, 21 | - | 80 ± 20 | 250 | 140 |
| Device B | 1.38, 1.10 | 1.32, 0.95 | 43, 22 | - | 45 ± 10 | 250 | 150 |

[a] Maximum and minimum values are displayed for each item when 75 nm misalignment is included between the MM and QDs in both lateral directions. Larger and smaller $b_{sl}$ are given for both QDs (note that $b_{sl}$ is larger at one QD than at the other). Interdot distance of 150 nm and the magnetization of Co ferromagnet of 1.8 T are assumed throughout this work.

[b] $t_{MM}$ is the thickness of the Co MM.

[c] $d_{2DEG}$ is the distance between the MM and the 2DEG constituting QDs, *i.e.* the sum of the depth of the 2DEG from the wafer surface and the insulator thickness.

However, we argue that fast ESR, with spin-flip frequencies sufficiently high for the demonstration of nontrivial two-spin operations in GaAs QDs as a benchmark platform [2,3], will require improvement of the MM field properties in the following ways. First, $b_{sl}$ needs to be ≳ 0.8 mT/nm to further enhance ESR rotation speed. Currently, ESR is the most time-consuming operation in the universal gate set. Given this field gradient and assuming typical experimental parameters in GaAs QDs [12], $f_{Rabi}$ of ≳ 50 MHz is accessible. This allows several spin rotations within the ensemble phase coherence (or dephasing) time $T_2^*$ of a few tens of ns in GaAs QDs [8,22], with improved gate-fidelity.

Secondly, for fast ESR, $\Delta B_Z$ has to be increased in proportion with $f_{Rabi}$. The reason is that to address single spins independently, the ESR peak separation $|g|\mu_B \Delta B_Z$ ($g$ is the Landé factor and $\mu_B$ is the Bohr magneton) must be greater than the ESR peak width $\Delta f \sim f_{Rabi}+T_2^{*-1}$ [8,23]. While for relatively slow ESR $\Delta f$ is dominated by the fluctuation of the Larmor frequency, mostly due to the Overhauser field, the contribution of $f_{Rabi}$ to $\Delta f$ becomes no longer negligible when the spin flip rate exceeds the dephasing rate, *i.e.* $f_{Rabi} > T_2^{*-1}$. Assuming $f_{Rabi}$ is more or less the same in all QDs, fast addressable ESR is only feasible when $|g|\mu_B \Delta B_Z > 2f_{Rabi}$. That is, for $f_{Rabi}$ = 50 MHz in GaAs QDs with $|g|$ = 0.40, $\Delta B_Z$ has to be > 18 mT.

Third, MM field properties should be tolerant of the relative misalignment between MM and QDs. Such misalignment is usually present in real QD devices as it can arise from overlay fabrication errors and inaccurate estimation of QD positions (QDs may not be formed exactly at positions expected from gate geometry, due to imperfect simulation and/or dopants in semiconductor wafers). It is difficult to reduce this error distance $d_{err}$, especially for a multi-qubit system. In reality, $d_{err}$ is typically 50-100 nm even with state-of-the-art semiconductor processing technology. Not surprisingly, this amount of misalignment can spoil the MM properties (Table I), since the stray field of the MM tends to be strongly position dependent.

In the following, we optimize the MM design relying on the simulation to meet all the following requirements clarified above: (1) $b_{sl}$ ≳ 0.8 mT/nm, (2) $\Delta B_Z$ > 18 mT, (3) misalignment robustness, *i.e.* conditions (1) and (2) are met in the presence of $d_{err}$ = 75 nm MM misalignment.

In general, the MM designs for QD devices can be categorized into two types: a "single" and a "paired" MM. A single MM can be specified by the length $l(x)$ in the magnetization axis $y$, where the $x$-axis is in the 2DEG plane and orthogonal to the $y$-axis (see Fig. 1(a)). A paired MM (see Fig. 1(b) and (c)), on the other hand, can be characterized by the gap width $g(x)$. If $l(x) = g(x)$, these different types of MMs are almost equivalent in that they produce in principle roughly the same field properties away from the MM edges, with the stray field directing in the opposite direction. Therefore, although in what follows we will restrict ourselves to the paired MM, the same type of reasoning will hold for the single MM as well.

First we examine how to realize a large $b_{sl}$ in the presence of finite misalignment. We distinguish between two main configurations of the QDs and MM, namely the "parallel" and the "perpendicular" one, which relate to the angle of the QD alignment axis with respect to the MM magnetization axis (// $B_0$ and // $y$). $b_{sl}$ becomes largest when we minimize displacements of QDs from the MM gap center ($y = 0$) and this is realized when QDs are on the MM center line. Therefore, the perpendicular configuration (see Fig. 1(b)) is more favorable than the parallel one to make $b_{sl}$ more robust against the misalignment. We note, however, that in the parallel configuration (see Fig. 1(a) and (c)), where the QD axis is aligned with the magnetization axis, a large $\Delta B_Z$ can be readily obtained with asymmetrical MMs.

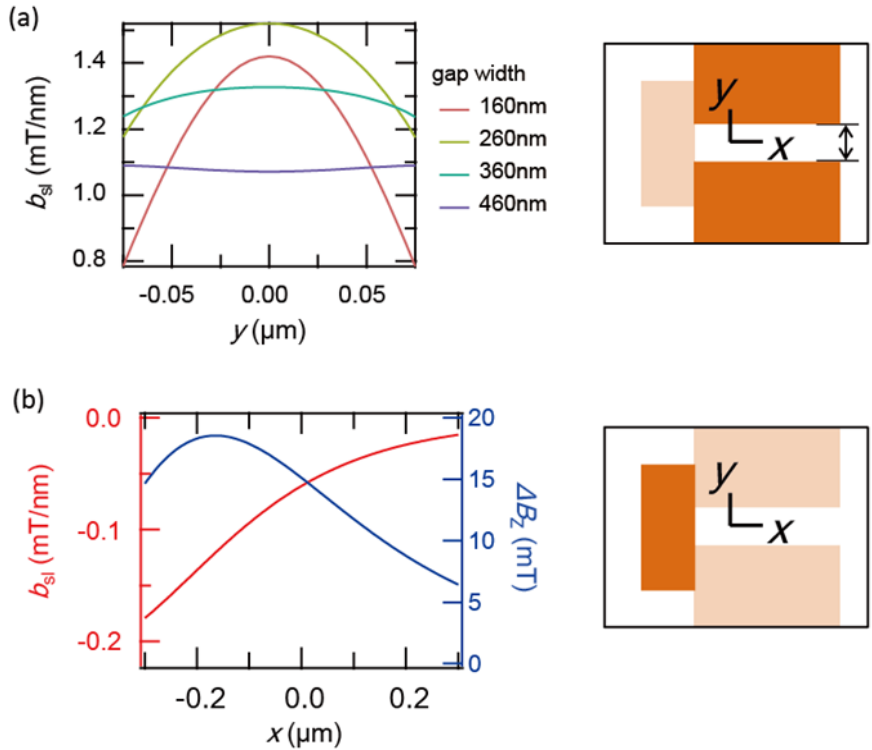

Figure 2. (a) Simulated slanting magnetic field for a pair of MMs with a constant gap in between. Relevant MMs are highlighted in deep orange in the inset. The magnetization axis is in the $y$ direction. Geometrical parameters (such as $t_{MM}$ and $d_{2DEG}$) are shown in Table I. $b_{sl}$ at $y = \pm 75$nm is maximized with the gap width of ~ 260 nm. (b) Simulated field properties considering the bridge part alone. The relevant part is highlighted in deep orange in the inset. In the range of $|x| < 75$ nm, $|b_{sl}|$ decays < 0.1 mT/nm (= 13 % of the required value) while $\Delta B_Z > 10$ mT (= 56 % of the required value).

Next we discuss how a good balance between a large $b_{sl}$ and a large $\Delta B_Z$ is achieved. A conventional method to produce $\Delta B_Z$ in the perpendicular configuration (as in Fig. 1(b)) is to taper the MMs. However, the simulation shows that this can harm the robustness of $b_{sl}$. The reason is related to the size of the gap between the paired MMs. When the gap becomes smaller than some optimal value for a given $d_{err}$, $b_{sl}$ gets more susceptive to the misalignment, whereas a larger gap will weaken $b_{sl}$ at the center (see Fig. 2(a)). For the robustness of $b_{sl}$, it is therefore important not to change $g(y)$ too far from this optimized value, which maximizes $b_{sl}$ with finite misalignment. This value is ~ 260 nm for the geometrical parameters such as $t_{MM}$ and $d_{2DEG}$ specified in Table I. Therefore, $g(y)$ should not be tapered much, but the problem with a paired MM with $g(y)$ roughly fixed to some value is that it can produce only a small $\Delta B_Z$.

We find through the simulation that a "bridge" structure is more favorable than the tapered structure to obtain the necessary $\Delta B_Z$ (see Fig. 2 (b) inset). The bridge part creates the Zeeman field (and also the slanting field) opposite to that induced by the rest (the paired MM with a constant gap). This change in sign brings about an abrupt distribution of Zeeman field and hence a large $\Delta B_Z$ among the QDs. Since $\Delta B_Z$ decays slower (see Fig. 2(b)) than $b_{sl}$, a sufficiently large $\Delta B_Z$ can be achieved with only a slight decrease of $b_{sl}$ using this approach. The advantage of this structure is in that unlike the tapered one the misalignment-robustness of $b_{sl}$ is ensured by the paired MM with an optimal, constant opening.

We note that by putting the bridge closer to the QDs, this structure can also be utilized to supply $\Delta B_Z$ exceeding 50 mT. This makes a single-step controlled-phase gate operating at ~ 20 MHz within experimental reach in GaAs QDs [3,24], and may help to construct efficient two-qubit gates other than CNOT and $\sqrt{SWAP}$.

An example of the optimized, bridged MM design is shown in Fig. 3(a). The constant gap of the split-pair part is chosen to maximize the minimum $b_{sl}$ within 75 nm from the MM axis, so that $b_{sl}$ becomes misalignment-proof by design. The position of the bridge is chosen to keep $\Delta B_Z > 18$ mT in the presence of a 75 nm misalignment. Figures 3(c) and (d) show the simulation results, where $b_{sl} > 0.9$ mT/nm for both QDs and $\Delta B_Z > 19$ mT in a 150 nm × 150 nm area. We would like to point out that the bridge design is less susceptible to misalignment than previous presented designs (Table I). To demonstrate the effectiveness of the presented design scheme, we incorporated this design of MM with a DQD device (device A). The ESR spectra measured in device A (Fig. 3(e)) shows $\Delta B_Z$ to be around 70 mT. This indicates that the average Zeeman-field gradient across the DQD is as large as 0.35 mT/nm (the simulated value is 0.14 - 0.32 mT/nm, see Table I), even when we assume a slightly large interdot distance of 200 nm to account for a relatively weak tunnel coupling observed in this device. This also ensures an ESR addressability for $f_{Rabi} > 100$ MHz.

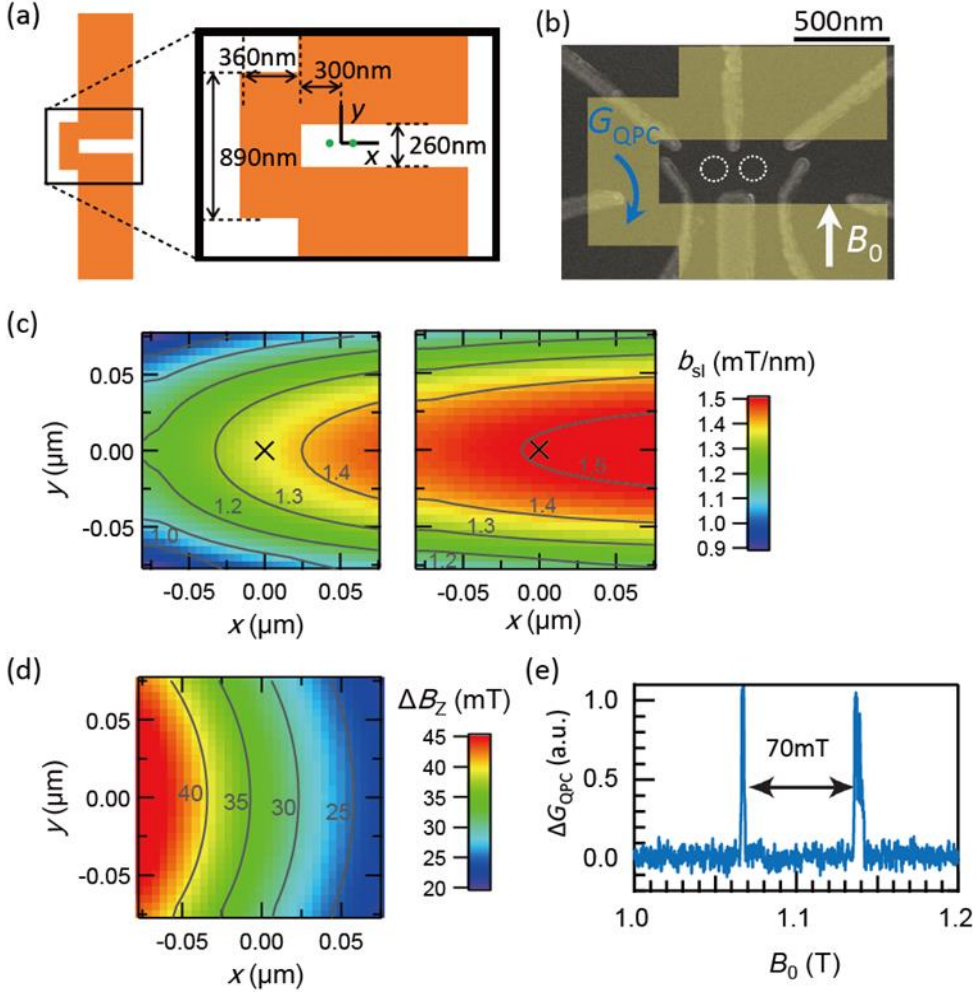

Figure 3. QD and MM in this work (Device A). (a) Robust MM design. QD positions are depicted as dots around $(x, y)=(0, 0)$. (b) Scanning electron micrograph. A modulation-doped GaAs/AlGaAs wafer with a 2DEG 57 nm below the surface was used for the fabrication. On top of the 100 nm insulating layer, a 250 nm thick Co MM was deposited. (c) Simulated $b_{sl}$ distribution for the left and right dot allowing for a 75nm misalignment. The crosses at the centers mark the expected QD positions without misalignment. Geometrical parameters used for the simulation are specified in Table I. (d) Simulated $\Delta B_Z$ distribution for the same parameters. Here $\Delta B_Z(x, y) = B_Z(x + 75\text{nm}, y) - B_Z(x - 75\text{nm}, y)$, where $B_Z(x, y)$ denotes the local Zeeman field at $(x, y)$. (e) Measured ESR spectra. Broader peak widths indicate improved ESR Rabi frequencies above the nuclear dephasing rates [23].

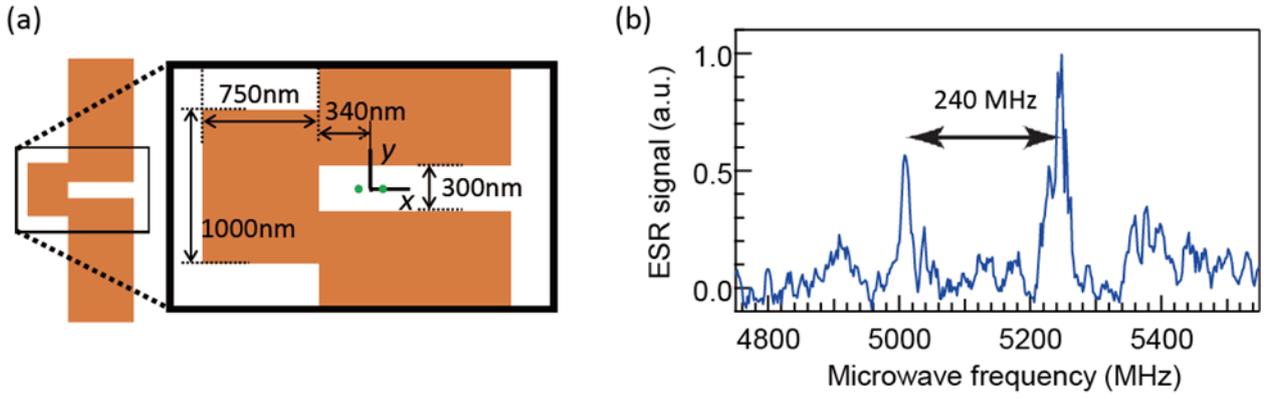

Figure 4. Device B. (a) MM designed for a different geometrical parameter than Device A. (b) Measured ESR spectra, with a peak separation of 240 MHz. Given the measured value of $|g|$ = 0.36, this corresponds to $\Delta B_Z$ = 48 mT.

For device B we also used the bridge design. Here, a peak separation $\Delta B_Z$ of about 60 mT was experimentally observed. Again, excellent agreement between experiment and simulation is shown (Fig. 4(b) and Table I).

In summary, we discussed crucial requirements of the MM stray field to obtain fast spin rotation in QDs using MMs *i.e.* a misalignment robust $b_{sl}$ and $\Delta B_Z$. We proposed a new design to satisfy these requirements by employing a constant gap and a bridge in a paired Co MM. The proposed design gives $b_{sl}$ > 0.9 mT/nm and $\Delta B_Z$ > 19 mT for realistic device parameters up to a misalignment of 75 nm of the MM. This facilitates the realization of fast addressable ESR ($\gtrsim$ 50 MHz in GaAs QDs) [8] as well as other quantum gate operations involving ESR such as the CNOT gate [3]. The MM field properties can be even further enhanced by decreasing the distance between the MM and the QDs as well as using a MM material with stronger magnetization than Co. Since the spin-operation scheme is material-independent, the proposed design would be useful for optimizing the performance of nanostructure-based spintronic devices and quantum information processors and for precise control of a solid-state electron spin and its superpositions in various materials.

**Acknowledgements**

We thank G. Allison for carefully reading the manuscript. Part of this work is financially supported by ImPACT Program of Council for Science, Technology and Innovation (Cabinet Office, Government of Japan), Project for Developing Innovation Systems from the Ministry of Education, Culture, Sports, Science and Technology, Japan (MEXT), Grant-in-Aid for Scientific

Research(S) (No. 26220710) from Japan Society for the Promotion of Science (JSPS), and the IARPA project "Multi-Qubit Coherent Operations" through Copenhagen University. T.O. acknowledges financial support from the Grant-in-Aid for Research Young Scientists B, Yazaki Memorial Foundation for Science and Technology Research Grant, Japan Prize Foundation Research Grant, Advanced Technology Institute Research Grant, Toyota Physical & Chemical Research Institute Scholars, and RIKEN Incentive Research Project. M.P.-L. acknowledges financial support from the Natural Science and Engineering Research Council of Canada (NSERC). J.Y. and T.T. acknowledge the financial support from JSPS for the Research Fellowships. R.B. gratefully acknowledges financial support by the Austrian Federal Government (in particular from Bundesministerium für Verkehr, Innovation und Technologie and Bundesministerium für Wissenschaft, Forschung und Wirtschaft) represented by Österreichische Forschungsförderungsgesellschaft mbH and the Styrian and the Tyrolean Provincial Government, represented by Steirische Wirtschaftsförderungsgesellschaft mbH and Standortagentur Tirol, within the framework of the COMET Funding Programme.

## References

[1] I. Žutić, J. Fabian, and S. Das Sarma, Reviews of Modern Physics 76, 323 (2004).

[2] D. D. Awschalom, L. C. Bassett, A. S. Dzurak, E. L. Hu, and J. R. Petta, Science **339** (6124), 1174 (2013).

[3] D. Loss and D. P. DiVincenzo, Physical Review A **57** (1), 120 (1998).

[4] F. H. Koppens, C. Buizert, K. J. Tielrooij, I. T. Vink, K. C. Nowack, T. Meunier, L. P. Kouwenhoven, and L. M. Vandersypen, Nature **442** (7104), 766 (2006).

[5] K. C. Nowack, F. H. Koppens, Y. V. Nazarov, and L. M. Vandersypen, Science **318** (5855), 1430 (2007).

[6] E. A. Laird, C. Barthel, E. I. Rashba, C. M. Marcus, M. P. Hanson and A. C. Gossard, Phys. Rev. Lett. **99**, 246601 (2007).

[7] M. Pioro-Ladrière, T. Obata, Y. Tokura, Y.-S. Shin, T. Kubo, K. Yoshida, T. Taniyama, and S. Tarucha, Nature Physics **4**, 776 (2008).

[8] J. Yoneda, T. Otsuka, T. Nakajima, T. Takakura, T. Obata, M. Pioro-Ladrière, H. Lu, C. J. Palmstrøm, A. C. Gossard, and S. Tarucha, Phys. Rev. Lett. **113**, 267601 (2014).

[9] J. R. Petta, A. C. Johnson, J. M. Taylor, E. A. Laird, A. Yacoby, M. D. Lukin, C. M. Marcus, M. P. Hanson, and A. C. Gossard, Science **309** (5744), 2180 (2005).

[10] R. Brunner, Y.-S. Shin, T. Obata, M. Pioro-Ladrière, T. Kubo, K. Yoshida, T. Taniyama, Y. Tokura and S. Tarucha, Phys. Rev. Lett. **107**, 146801 (2011).

[11] Y. Tokura, W. G. van der Wiel, T. Obata, and S. Tarucha, Phys. Rev. Lett. **96**, 047202 (2006).

[12] T. Obata, M. Pioro-Ladrière, Y. Tokura, Y.-S. Shin, T. Kubo, K. Yoshida, T. Taniyama and S. Tarucha, Phys. Rev. B **81**, 085317 (2010).

[13] E. Kawakami, P. Scarlino, D. R. Ward, F. R. Braakman, D. E. Savage, M. G. Lagally, M. Friesen, S. N. Coppersmith, M. A. Eriksson, and L. M. K. Vandersypen, Nature

nanotechnology **9**, 666 (2014).

[14] X. Wu, D. R. Ward, J. R. Prance, D. Kim, J. K. Gamble, R. T. Mohr, Z. Shi, D. E. Savage, M. G. Lagally, M. Friesen, S. N. Coppersmith, and M. A. Eriksson, Proceedings of the National Academy of Sciences **111** (33), 11938 (2014).

[15] T. Takakura, M. Pioro-Ladrière, T. Obata, Y.-S. Shin, R. Brunner, K. Yoshida, T. Taniyama and S. Tarucha, Appl. Phys. Lett. **97**, 212104 (2010).

[16] V. N. Golovach, M. Borhani, and D. Loss, Physical Review B **74**, 165319 (2006).

[17] T. Obata et al. unpublished.

[18] In MM-ESR, the Rabi frequency can be written as $f_{\text{Rabi}} = |g|\mu_B e E_{\text{ac}} l_{\text{orb}}^2 b_{\text{sl}}/(2h\Delta)$, where $g$ is the Landé factor and $\mu_B$ is the Bohr magneton, $e$ is a single electron charge, $l_{\text{orb}}$ is the orbital spread, $h$ is Planck's constant and $\Delta$ is the QD confinement energy [11,7]. In GaAs QDs, typical values for these are $g = -0.40$, $E_{\text{ac}} = 5$ mV/μm, $l_{\text{orb}} = 50$ nm and $\Delta = 0.5$ meV [12,5]. The upper bound on $E_{\text{ac}}$ in the experiment is set by photon-assisted-tunneling obscuring ESR signals and may be improved by refining QD gate patterns, the frequency applied and so on.

[19] We used MATHEMATICA RADIA package available at http://www.esrf.fr/ for MM field calculations.

[20] K. Ono, D. G. Austing, Y. Tokura, and S. Tarucha, Science **297** (5585), 1313 (2002).

[21] F. H. Koppens, J. A. Folk, J. M. Elzerman, R. Hanson, L. H. van Beveren, I. T. Vink, H. P. Tranitz, W. Wegscheider, L. P. Kouwenhoven, and L. M. Vandersypen, Science **309** (5739), 1346 (2005).

[22] F.H.L. Koppens, K.C. Nowack, and L.M.K. Vandersypen, Phys. Rev. Lett. **100**, 236802 (2008).

[23] The expression for $\Delta f$ can be obtained, based on a rotating wave approximation under static random nuclear field, as a Voigt linewidth. This is a result of the Lorentzian profile due to power broadening and the Gaussian profile due to inhomogeneous broadening. The Voigt peak width can be approximated as $\Delta f \approx c_1 f_{\text{Rabi}} + \sqrt{c_2 f_{\text{Rabi}}^2 + (\alpha T_2^*)^{-2}}$ (Ref. [25]) with $c_1 \doteq 0.535$ and $c_2 \doteq 0.215$ and $\alpha = \pi/(2\sqrt{\ln 2})$. The fact that $\Delta f > f_{\text{Rabi}}$ (due to power broadening) can also be viewed as a consequence of the time-frequency uncertainty principle that states that the product of the operation time and spectral bandwidth must be greater than $h/2\pi$.

[24] T. Meunier, V. E. Calado, and L. M. K. Vandersypen, Phys. Rev. B **83**, 121403(R) (2011).

[25] J. J. Olivero and R. L. Longbothum, Journal of Quantitative Spectroscopy and Radiative Transfer **17** (2), 233 (1977).